\documentstyle[12pt,psfig]{article}
\def\lsim{\:\raisebox{-0.5ex}{$\stackrel{\textstyle<}{\sim}$}\:}
\def\gsim{\:\raisebox{-0.5ex}{$\stackrel{\textstyle>}{\sim}$}\:}
\def\fig#1{{Fig. (\ref{#1})}}

\def\21{$SU(2) \ot U(1)$}
\def\321{$SU(3) \ot SU(2) \ot U(1)$}

{\bf

\def\n.c.#1#2#3{         { Nuovo Cim. }{\bf #1}, #3 (19#2)}
\def\r.n.c.#1#2#3{       { Riv. del Nuovo Cim. }{\bf #1}, #3 (19#2)}

\begin{document}
\thispagestyle{empty}
\begin{titlepage}
\vskip 0.3cm
\begin{center}
{\bf \large
Testing for New Physics with Low-Energy Anti-Neutrino Sources: 
LAMA as a Case Study
}
\end{center}
\normalsize
\begin{center}
{\bf I. Barabanov~$^a$}, 
{\bf P. Belli~$^b$},
{\bf R. Bernabei~$^b$},
{\bf V.I.Gurentsov~$^a$, V.N.Kornoukhov~$^a$},
{\bf O. G. Miranda~$^c$}\footnote{ E-mail: omr@flamenco.ific.uv.es
On leave from {\sl Departamento de F\'{\i}sica CINVESTAV-IPN, A. P. 
14-740, M\'exico 07000, D. F., M\'exico.}}
{\bf V. B. Semikoz~$^d$},
{\bf and J.~W.~F. Valle~$^c$}\footnote{ E-mail valle@flamenco.ific.uv.es}\\
\end{center}
\begin{center}
{\it $^a$ Institute for Nuclear Research, Moscow, Russia}
\end{center}
\begin{center}
{\it $^b$ Dip. di Fisica Universita' di Roma ``Tor Vergata'', \\ 
and INFN, sez. Roma 2, I-00133 Rome, Italy} \\
\end{center}
\begin{center}
{\it $^c$ Instituto de F\'{\i}sica Corpuscular - C.S.I.C.\\
Departament de F\'{\i}sica Te\`orica, Universitat de Val\`encia\\}
{\it 46100 Burjassot, Val\`encia, SPAIN \\
http://neutrinos.uv.es        }\\
\end{center}
\begin{center}
{\it $^d$ The Institute of the Terrestrial Magnetism, \\ 
the Ionosphere and Radio Wave Propagation of the Russian Academy of Science, \\
IZMIRAN, Troitsk, Moscow region, 142092, Russia} \\
\end{center}
\baselineskip=13pt

\begin{abstract}

Some electroweak models with extended neutral currents, such as those
based on the $E_6$ group, lead to an increase of the
$\overline{\nu}-e$ scattering cross section at energies below 100 keV.
We propose to search for the heavy $Z^\prime$ boson contribution in an
experiment with a high-activity artificial neutrino source and with a
large-mass detector. We present the case for the LAMA experiment with
a large NaI(Tl) detector located at the Gran Sasso underground
laboratory. The neutrino flux is known to within a one percent
accuracy, in contrast to the reactor case and one can reach lower
neutrino energies. Both features make our proposed experiment more
sensitive to extended gauge models, such as the $\chi$ model. For a
low enough background the sensitivity to the $Z_\chi$ boson mass would
reach 600 GeV for one year running of the experiment.

pacs{13.15.+g  12.20.Fv  12.60.Cn  95.55.Vj }
\end{abstract}
\end{titlepage}
\vskip 1cm

\section{Introduction}

We have recently revived the idea of using low energy neutrino
reactions as sensitive precision tests of the Standard Model (SM)
\cite{Barabanov} both as probes for non-standard neutrino electromagnetic
properties \cite{Vogel,Ferrari} as well as extended gauge structures
\cite{msv}. The alternative use of reactors such as the MUNU
experiment has recently been suggested \cite{Broggini}.  Here we
discuss in detail the possible advantages of performing measurements
at low energy with a very intense anti-neutrino source, such as a
$^{147}Pm$ source proposed in the LAMA experiment \cite{Barabanov}.
In contrast to ref.  \cite{msv} where the potentiality of low energy
source experiments was discussed from a more general perspective, here
we concentrate on the specific case of the LAMA proposal, using the
design parameters of the experiment as a case study.  In LAMA the
neutrino flux should be determined with an accuracy better than one
percent. The dimensions of the source will allow to surround it by the
detector.

We evaluate the potential that such an isotope source offers in
testing the electroweak gauge structure of the SM. For definiteness we
consider the sensitivity of this experiment to models that can arise
from an underlying $E_6$ framework \cite{fae}. The later have been
quite popular since the eighties, especially because they arise in a
class of heterotic string compactifications. LEP measurements at the Z
peak have achieved very high precision in determining the neutral
current coupling constants governing $e^{+}e^{-} \to l^{+}l^{-}$. This
constrains especially the mixing angle between the $Z$ and the
$Z^\prime$ bosons. As a result we will focus here on the possible
constraint for the mass of the additional gauge boson
$M_{Z^\prime}$. 

We consider extensions of the Standard Model involving an extra $U(1)$
symmetry at low-energies, coupled to the following hyper-charge \cite{fae}
\begin{equation}
Y_\beta =\mbox{cos}\beta Y_{\chi}+\mbox{sin} \beta Y_{\psi}, 
\end{equation}
while the charge operator is given as $ Q=T^3+ Y$.  The values of
$Y_{\chi}$ and $Y_{\psi}$ for different particles are well known in
the literature and are given in Table 1. Any value of $\beta$ is
allowed, giving us a continuum spectrum of possible models of the weak
interaction. The most common choices considered in the literature are
$\mbox{cos} \beta =1$ ($\chi$ model), $\mbox{cos} \beta =0$ ($\psi$
model) and $\mbox{cos} \beta =\frac{\sqrt{3}}{\sqrt{8}}$, $\mbox{sin}
\beta =-\frac{\sqrt{5}}{\sqrt{8}}$ ($\eta$ model).
The masses of the neutral gauge bosons in these models arise from a
$2\times 2$ matrix which may always be written as
\begin{equation}
\pmatrix{M^2_{Z^0} & \mu^2 \cr \mu^2 & M^2 }
\label{matrix}
\end{equation}
where $M^2_{Z^0}$ would be the Z-mass in the absence of mixing with
the extra gauge boson. The eigenstates following from
Eq. (\ref{matrix}) are given by
\begin{eqnarray}
Z&=& cos \theta^\prime Z_0 - sin\theta^\prime Z^\prime_0 \nonumber \\
Z^\prime&=& sin\theta^\prime Z_0 + cos\theta^\prime Z^\prime_0 
\label{mixing}
\end{eqnarray}
in terms of the weak eigenstate gauge fields $Z_0$ and $Z^\prime_0$.

If we restrict ourselves to the case when only doublet and singlet
Higgs bosons arising from the fundamental {\bf 27}-dimensional
representation of the primordial $E_6$ group are present, then we have
the following expression for the mixing parameter \cite{GCprd}
\begin{equation}
\mu^2 = M^2_{Z_0} sin\theta_{W} \big[ \frac13 
\sqrt{10}(1-2\xi)sin\beta-\sqrt{2/3}cos\beta \big] , \label{cons}
\end{equation}
depending on the chosen model through the angle $\beta$ and also
through the parameter $\xi$, defined as $\xi =
\frac{v^2_d}{v^2_u+v^2_d}$, where $v_u$ and $v_d$ are the vacuum
expectation values dominantly responsible for the electroweak
breaking.  Such models were called {\sl constrained superstring
models} in ref. \cite{npb345}.  For the $\chi$ model we do not have
any dependence on the $\xi$ parameter, since in this case $sin\beta=0$
and therefore any $\xi$ dependence in Eq. (\ref{cons}) is washed out.
As a result one obtains a direct relationship between the $Z^\prime$
mixing angle $\theta '$ and its mass $M_{Z^\prime}$.  This enables us
to convert any bound on $M_{Z^\prime}$ into a corresponding one on the
$Z^\prime$ mixing angle. The advantage of these models for us is that
they allow us to restrict the $Z^\prime$ mixing angle which is not obtained
directly by our present method.

\section{The Cross Section}

In the SM, the differential cross section for $\overline{\nu_{e}} e \to
\overline{\nu_{e}} e$ scattering  is given by,

\begin{equation}
\frac{d\sigma}{dT} = \frac{2m_{e}G^2_{F}}{\pi}\big\{ 
    (g_{L}+1)^2+ g_{R}^2 - 
    [2(g_{L}+1)^2+\frac{m_e}{E_{\nu}} (g_{L}+1)
    g_{R}]\frac{T}{E_{\nu}} + 
     (g_{L}+1)^2(\frac{T}{E_{\nu}})^2 \big\}  \label{DCS}
\end{equation}
where $g_L$ and $g_R$ are the Standard Model model expressions
$g_{L,R}=1/2(g_V \mp g_A)$. Here $T$ is the electron recoil energy and
$E_{\nu}$ the neutrino energy.

The neutral current contribution to $\overline{\nu} e \to
\overline{\nu} e$ scattering in extended models is given for example
in Ref. \cite{npb345}. The extra contribution due to the $Z^\prime$ to
the differential cross section will be, for $\theta '=0$,
\begin{eqnarray}
\delta\frac{d\sigma}{dT} = &\gamma &\frac{2m_{e}G^2_{F}}{\pi}\big\{ 
    2(g_L+1)\delta g_{L}+ 2g_R \delta g_{R} \nonumber \\
    & - [ & 4(g_L+1)\delta g_{L}+\frac{m_e}{E_{\nu}} 
    ((g_L+1)\delta g_{R}+
    g_R\delta g_{L})]\frac{T}{E_{\nu}} \nonumber \\
    &+ 2&(g_L+1)\delta g_{L}(\frac{T}{E_{\nu}})^2 \big\}  \label{dDCS}
\end{eqnarray}
where
\begin{equation}
\gamma = \frac{M^2_{Z}}{M^2_{Z^\prime}}
\end{equation}
and $\delta g_{L,R}$ depend on the model under consideration.  For the
particular case of the LRSM \cite{LR1,LR2} this corrections are given by

\begin{eqnarray}
\delta g_{L} &=& \frac{s^4_{W}}{r^2_{W}}g_L 
                 +\frac{s^2_{W}c^2_{W}}{r^2_{W}}g_{R}  \\
\delta g_{R} &=&\frac{s^4_{W}}{r^2_{W}}g_R 
                 +\frac{s^2_{W}c^2_{W}}{r^2_{W}}g_{L} 
\end{eqnarray} 
where $s_{W}=sin\theta_{W}$, $c_{W}=cos\theta_{W}$ and
$r^2_{W}=cos2\theta_{W}$; while for the $E_6$ models we have, again
for $\theta '=0$,
\begin{eqnarray}
\delta g_{L} &=& 4\rho s^2_{W} (\frac{3\mbox{cos}\beta}{2\sqrt{24}}
        +\frac{\sqrt{5}}{\sqrt{8}}\frac{\mbox{sin}\beta}{6})
        (\frac{3\mbox{cos}\beta}{\sqrt{24}}
         +\frac{\sqrt{5}}{\sqrt{8}}\mbox{sin} \beta)   \\
\delta g_{R} &=& 4\rho s^2_{W} (\frac{3\mbox{cos}\beta}{2\sqrt{24}}
        +\frac{\sqrt{5}}{\sqrt{8}}\frac{\mbox{sin}\beta}{6})
        (\frac{\mbox{cos}\beta}{\sqrt{24}}
         -\frac{\sqrt{5}}{\sqrt{8}}\mbox{sin}\beta) 
\end{eqnarray}
where, $\rho$ denotes the radiative corrections to the ratio
$M^2_{W}/M^2_{Z}cos^2\theta_{W} \equiv 1$ and $\beta$ defines the
$E_6$ model, in which we are interested in. All expressions we have
shown are for the case of $\theta '=0$ and in what follows we will
always use this assumption.

We can rewrite Eq. (\ref{dDCS}) as 
\begin{equation}
\delta\frac{d\sigma}{dT} =\gamma\Delta = \gamma\frac{2m_{e}G^2_{F}}{\pi}\big\{ 
    D  + E \frac{T}{E_{\nu}}(\frac{T}{E_{\nu}}-2) - 
    F \frac{m_e}{E_{\nu}} \frac{T}{E_{\nu}}\big\} \label{dc}
\end{equation}
with $\Delta$ in obvious notation and 
\begin{eqnarray}
D&=&2(g_L+1)\delta g_{L}+ 2g_R \delta g_{R} \\
E&=&2(g_L+1)\delta g_{L} \\
F&=&(g_L+1)\delta g_{R}+
    g_R\delta g_{L} .  \label{coe}
\end{eqnarray}

The correction to the $\overline \nu_{e} e$ scattering depends on the
model as well as on the energy region. In order to illustrate how
these corrections may affect the Standard Model prediction we can
define the expression

\begin{equation}
R=\frac{\Delta}{(\frac{d\sigma}{dT})^{SM}} \label{ratio}
\end{equation}

This ratio depends on the specific model through the angle $\beta$ and
depends also on the electron recoil energy as well as on the neutrino
energy.

\begin{figure}
\centerline{\protect\hbox{\psfig{file=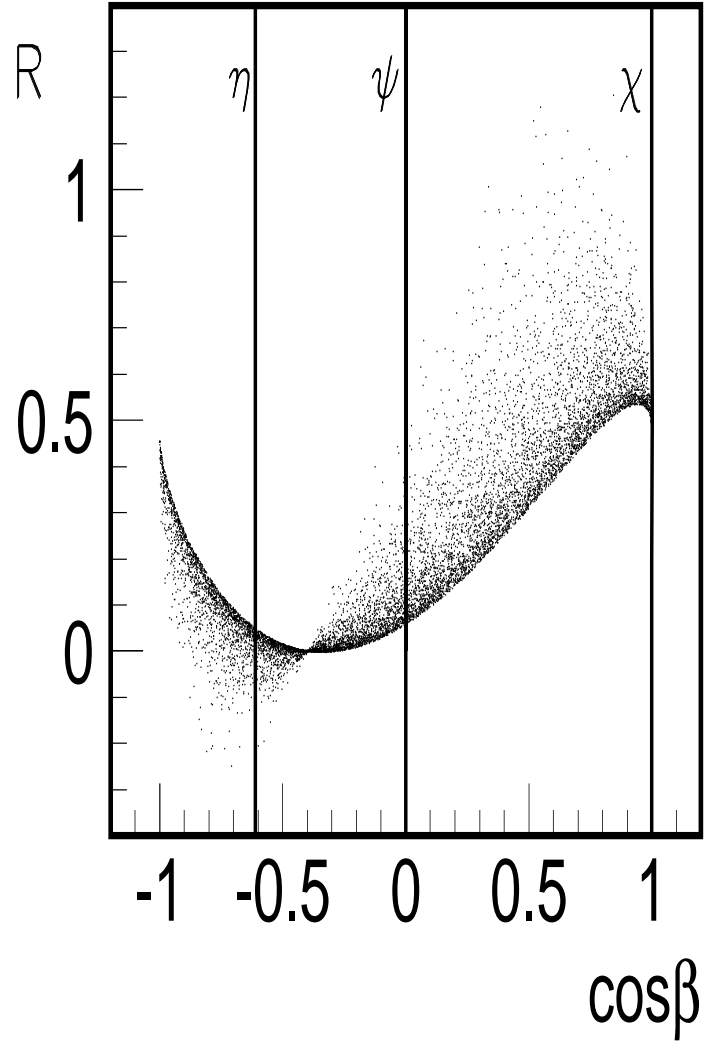,height=15cm,width=10cm}}}
\caption{Scatter plot of attainable values for the relative deviation 
R in eq.\protect{\ref{ratio}} for different values for the incoming
neutrino energy $E_{\nu}$ and the electron recoil energy $T$. The plot
is for different $E_6$ models. The $\chi$ model corresponds to $\cos
\beta =1$. Here we  assumed  $\theta ' =0$.}
\label{rat}
\end{figure}

We have plotted in \fig{rat} the quantity R in Eq. (\ref{ratio}) for a
class of $E_6$ models.  Different values of $E_{\nu}$ and $T$ are
considered, corresponding to the case of a $^{147}Pm$ source. We can
see from the plot that for $\mbox{cos}\beta \simeq 0.8$ one can have a
large deviation from the Standard Model predictions simply by varying
$E_{\nu}$ and $T$. For $\mbox{cos}\beta \simeq -0.4$ one can see that,
irrespective of the kinematical variables of $E_{\nu}$ and $T$ the
deviation that can be achieved is very small.  For models with
$\mbox{cos}\beta \lsim -0.4$ we have a negative contribution that
would decrease the number of events for some electron energies. This
effect is just the opposite of what would be expected in the case of a
neutrino magnetic moment and would be a direct signature of extended
gauge theories. However the sensitivity is smaller here than for the
models for which $\mbox{cos}\beta \simeq 0.8$.

Altogether, one sees that the $\chi$ model is the most sensitive to
this scattering. Other popular cases such as the $\eta$ \footnote{ For
simplicity of presentation we have chosen to plot this model as
corresponding to the values $\mbox{cos} \beta
=-\frac{\sqrt{3}}{\sqrt{8}}$ and $\mbox{sin} \beta
=+\frac{\sqrt{5}}{\sqrt{8}}$. We can do this since, as can be seen
from eq. (10) and (11) a simultaneous change in the signs of
$\mbox{sin} \beta$ and $\mbox{cos} \beta$ does not affect R.} and
$\psi$ models, often cited in the literature, have a smaller
sensitivity.  For this reason, from now on we fix on $\chi$ model,
which is also theoretically appealing as it corresponds to the
hypercharge that lies in $SO(10)/SU(5)$. For this case we wish to see
for which choices of the kinematical variables of $E_{\nu}$ and $T$
there is greater sensitivity to the new physics.

\begin{figure}
\centerline{\protect\hbox{\psfig{file=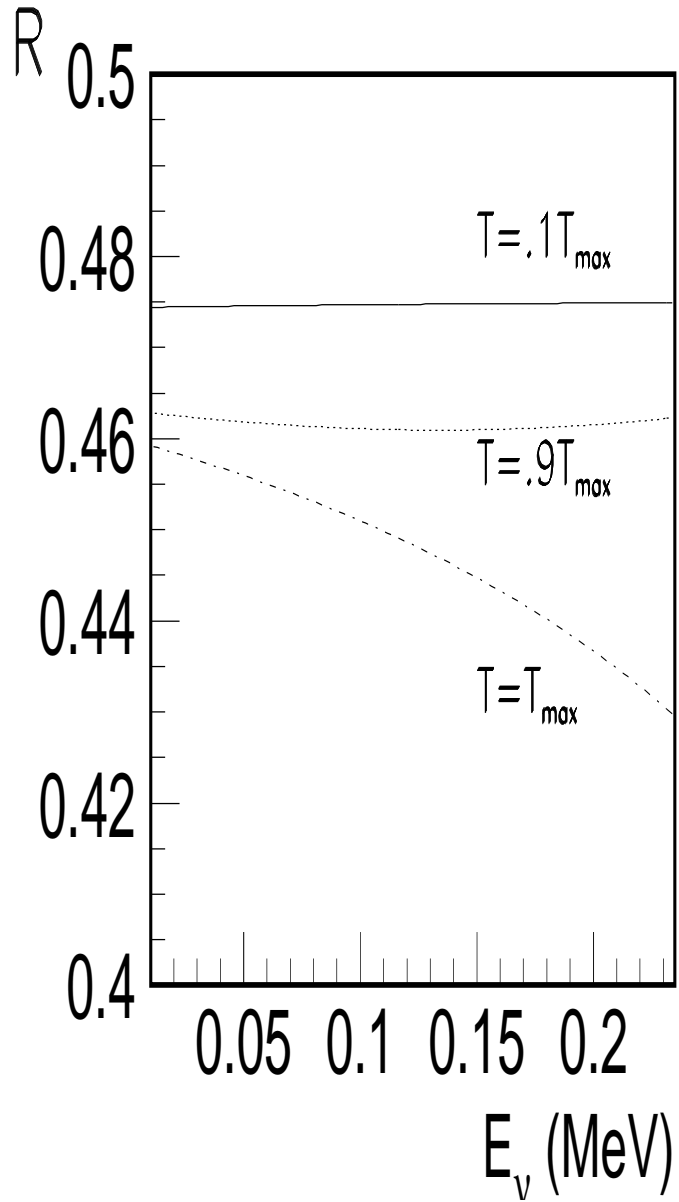,height=18cm,width=10cm}}}
\caption{Plot of the ratio given in eq.\protect{\ref{ratio}} 
for the $\chi$ model as a function of $E_{\nu}$ (in MeV). Different
values for the electron recoil energy, $T$, are shown.  Here we
assumed $\theta ' =0$.}
\label{ratchi}
\end{figure}

In \fig{ratchi} we have plotted Eq. (\ref{ratio}) for the specific
case of the $\chi$ model for the range of neutrino energies accessible
at a $^{147}Pm$ source.  In the plot we show the value of this ratio
for different values of the electron recoil energy.  For very low
electron energies, close to the energy threshold, one has a bigger
deviation from Standard Model predictions in the $\chi$ model. Indeed,
the independent term $D$ in eq.  (\ref{dc}) is large and positive,
while the other two terms give small negative contributions. The plot
shows how the corrections get smaller as the recoil electron energy
increases. As we can see from the figure, in order to reach a
constraint for $\gamma\lsim 0.1$, i.e. a $Z^\prime$ mass of about
$\gsim 300$ GeV or so, we need a resolution of the order of 5~\%.
From the present global fit of electroweak data the constraint on the
$Z^\prime$ mass for the $\chi$ model is 330 GeV at 95 \%
C. L. \cite{lang}, while from direct searches at the Tevatron the
constraint is 595 GeV at 95 \% C. L. \cite{CDF}.

In the next sections we will estimate the sensitivity of the LAMA
experiment to the $\chi$ model $Z^\prime$ boson.  This estimate is
obtained for the idealised case of anti-neutrino electron scattering
from a free electron. In practice in a realistic detector, such as
$NaI$, the electrons are bound and, in some cases it is necessary to
take the effect of binding into account, such as for the case of the
iodine atom.  Recently the corrections to the differential cross
section due to the atomic binding energy have been discussed in
ref. \cite{Fayans}. They are particularly important for the first
level of the $I$ atoms, where the binding energy is
$\varepsilon_1=32.92 $~KeV. These corrections could be taken into
account by computing the wave equation for the electron in the
Hartree-Fock-Dirac approximation \cite{Fayans2}. A useful
approximation can be obtained in terms of $q=\varepsilon_i + T$ where
$T$ is the recoil electron energy and $\varepsilon_i$ denotes the
electron binding energy in the $I$ atom \cite{Fayans2,Fayans3}. In
this case the energy distribution $S_{inel}(q)$ for the bound electron
of a given atomic level can be taken as $S_{inel}(q) \simeq
S_{free}(q)\theta (q-\varepsilon_i)$ where $S_{free}(q)$ stands for
the energy distribution in the free electron case.  We estimate an
overall uncertainty of about 10 \% in our use of the free electron
approximation, which would affect the overall statistics (expected
number of events in either the Standard Model or its extensions). Note
however that they should cancel in the ratio given in eq. \ref{ratio}
and in the corresponding sensitivity plots for the deviations from the
Standard Model that we have presented. 

\section{Experimental Prospects}

{\bf The LAMA experiment } 

The experiment LAMA \cite{Barabanov} has been proposed for neutrino
magnetic moment search in the range of $10^{-10} -10^{-11}$ $\mu_{B}$.
The principle of our proposed experiment is similar to the reactor
one. A large neutrino magnetic moment (LMM) would significantly
contribute to the neutrino electron scattering process. As an
alternative to the reactor idea, we propose the use of an artificial
neutrino source (ANS).  Our main aim in the experiment is the
investigation of low-energy neutrino-electron scattering as a test for
a possible deviation from the Standard Model prediction. For
definiteness we focus on the possibility of investigating the gauge
structure of the electroweak interaction as mentioned above.  The use
of an ANS has a number of essential advantages with respect to the
reactor for the experiments with low-energy neutrinos:

\begin{itemize}
\item        (a) the effective neutrino flux with the proposed ANS 
should be at least 10 times higher than in a typical reactor. This
could potentially be increased in the future;
\item        (b) the accuracy of source activity determination should be as 
high as a few tenths of percent in comparison with ~10 \% for the
reactor case;
\item        (c) the ANS energy should be low enough,  giving the 
possibility to minimise effect of background from the high-energy part of
the neutrino spectrum;
\item        (d) the experiment should  be carried out in a deep
underground laboratory, to ensure a low enough background.
\end{itemize} 

All these advantages make the use of an ANS preferable to a reactor
for the kind of experiment under consideration.

{\bf The isotope for the ANS  }

\begin{figure}
\centerline{\protect\hbox{\psfig{file=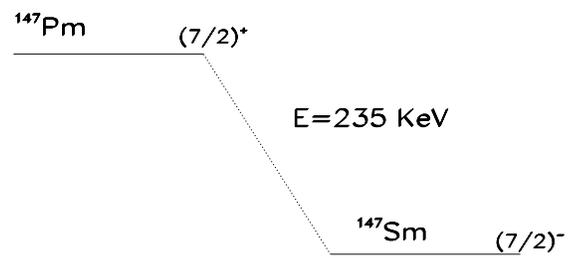,height=5cm,width=8cm}}}
\caption{Level scheme of $^{147}Pm$ nucleus.}
\label{level}
\end{figure}

The $^{147}$~Pm isotope has been chosen as an optimal candidate for
the ANS both from the point of view of its physical parameters (low
enough neutrino energy, absence of gamma-rays, long enough lifetime)
as well as the possibility to produce a large enough activity. The
scheme of the $^{147}$~Pm decay with its basic parameters is presented
in Fig \ref{level}.

The $^{147}$~Pm isotope has been produced commercially since 1980 by a
Russian Nuclear Plant called "Mayak" by extracting it from used reactor
fuel. The other radioactive REE elements admixture in the produced
$^{147}$~Pm is less then $10^{-9}$ and could be lower, if necessary.
A 5 MCi $^{147}$~Pm source is planned to be used for the first step of
the experiment. A source with such an activity can be produced by the
plant in 3-4 moths after small improvement in technology. An upgrade
in activity up to 15 Mci can be achieved by the plant in a reasonable
period of time. The neutrino spectrum of $^{147}$~Pm is well-known
from the experimental measurement of the beta-electron spectrum. It
corresponds to an allowed transition and is given in Fig 1 of
ref.~\cite{Kornoukhov}. We will use in the following the 5 MCi
$^{147}$~Pm source activity, unless mentioned otherwise. It is
important to notice that the neutrino flux from the source is expected
to be known with high accuracy (~0,3\% ) \cite{Kornoukhov}, thus
substantially reducing the systematic errors in contrast to the
recently suggested \cite{Broggini} reactor neutrino experiment.

{\bf Scheme of the experiment} 

As a target we consider the use of a NaI(Tl) scintillator installed at
the Gran Sasso Laboratory for dark matter particles search
\cite{Bernabei}. The NaI(Tl) detector is an ideal detector for this
kind of experiment, as the mass of the detector could be large enough,
a threshold of about 2 keV can be achieved \cite{Bernabei} and methods
of its purification from U, Th and K are well worked out and can be
further improved \cite{NCi98}.  The mass of detector is about 120 kg
and should be scaled up to 1 ton. For our estimate here we will use a
400 kg NaI mass as mentioned in ref.~\cite{Barabanov}.  The source is
surrounded by passive shielding of W (20 cm) and Cu (5 cm) from the
unavoidable admixture of $^{146}$~Pm ($\sim 10^{-8}$) with a 750 keV
$\gamma$ line and other possible REE gamma isotopes.  After careful
measurement of gamma admixtures in the commercial $^{147}$~Pm samples
the possibility to decrease the shielding will also be considered.
        
{\bf The expected number of events} 

\begin{figure}
\centerline{\protect\hbox{\psfig{file=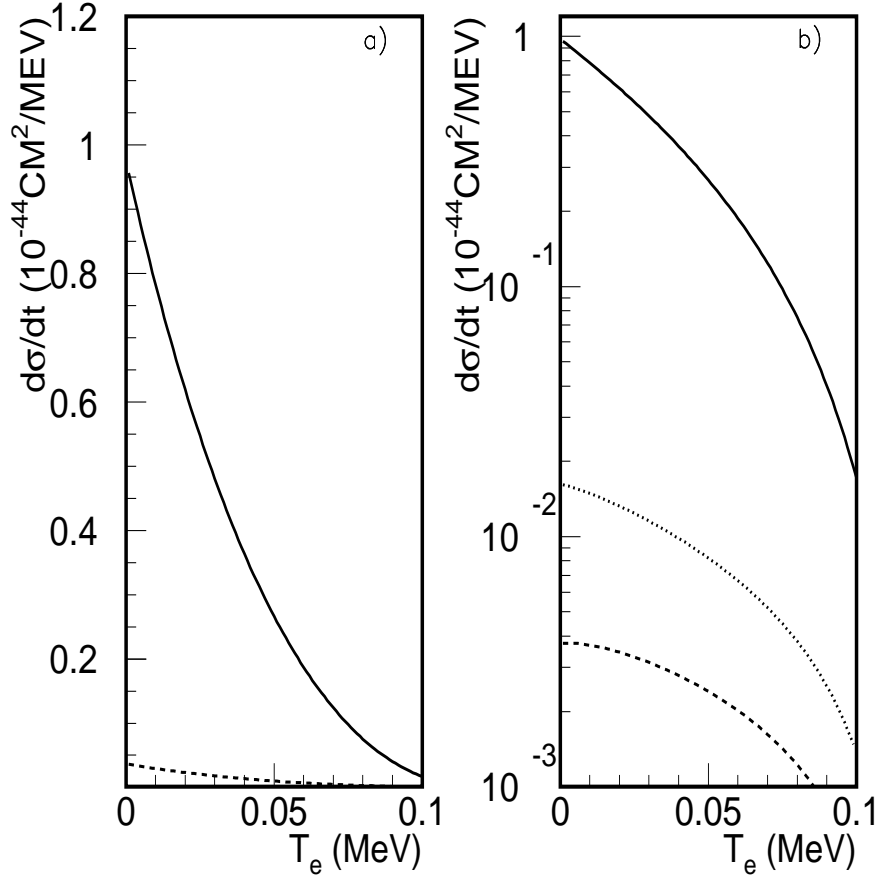,height=12cm,width=12cm}}}
\caption{Differential cross section for the $\overline{\nu_{e}} e$
scattering. In $a)$ we show the Standard Model case (solid line) and
the expected increment due to additional positive contribution of an
extra 330 GeV gauge boson in the $\chi$ model (dashed line). In figure
$b)$ we show the Standard Model case (solid line) and the expected
increment due to additional positive contribution from an extra 170
GeV gauge boson in the $\psi$ model (dotted line) and from and extra
390 GeV gauge boson in the left-right symmetric model (dashed
line). The differential cross section was integrated over the
antineutrino energy spectrum and averaged over the energy resolution.}
\label{cs}
\end{figure}

The basic cross section was given in Eq. (\ref{DCS}). This cross
section has to be integrated over the anti-neutrino energy spectrum
from the ANS and averaged over the expected detector energy
resolution. What we obtain for the case of interest ($^{147}$~Pm) is
shown in \fig{cs}.  In (a) we show the $\chi$ model, while in (b) we
give the results for the $\psi$ and LRSM models. As seen here the
effect for these models is rather small and we will not include them
in the following discussion.

\begin{figure}
\centerline{\protect\hbox{\psfig{file=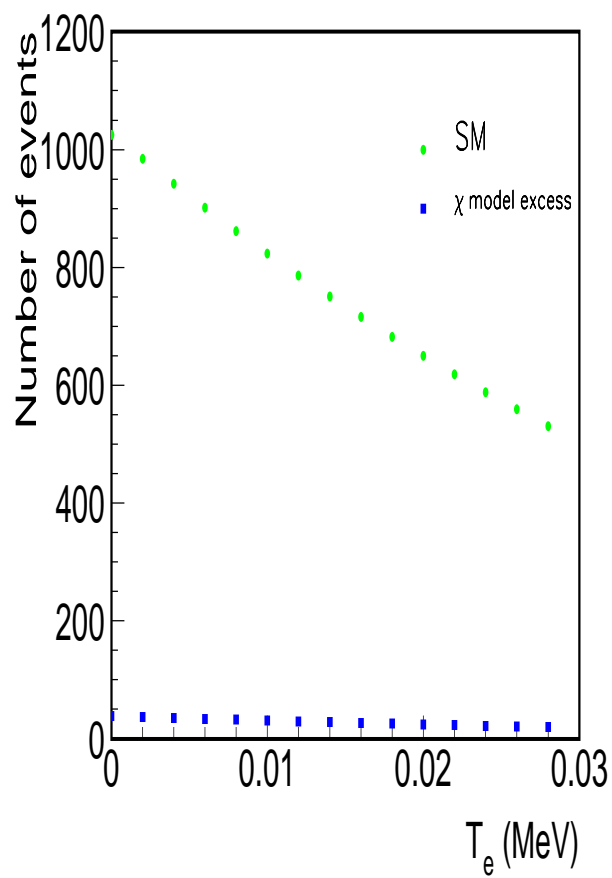,height=12cm,width=8cm}}}
\caption{Expected number of events per bin (2 KeV width) in the 
standard model and in the $\chi$ model for the parameters discussed in
the text.}
\label{his}
\end{figure}

The number of $\overline{\nu}-e$ scattering events expected in the
Standard Model and in the $\chi$ model are shown in Table 2. The
results in Table 2 assume a 400 kg detector and a 5 MCi $^{147}$~Pm
source with the geometry presented in Fig. 1 of Ref.~\cite{Barabanov}.
The total number of events is about $10^4$, ensuring a statistical
accuracy of about 1\%. The deviation from the Standard Model
prediction induced by an extra neutral gauge boson with mass of about
300 GeV is 4 \% or so, and could be detected.  As can be seen the
deviations are maximal for the lowest energies. It is therefore
convenient to present the results for a region up to 30 KeV. In
Fig \ref{his} we display the results we obtain when restricting to
this optimum region. The plot shows the histogram obtained when we bin
the recoil energy variable in 2 KeV bins, the increment due to a 330
GeV neutral gauge boson in the $\chi$ model is also shown at the
bottom of the figure.

\begin{figure}
\centerline{\protect\hbox{\psfig{file=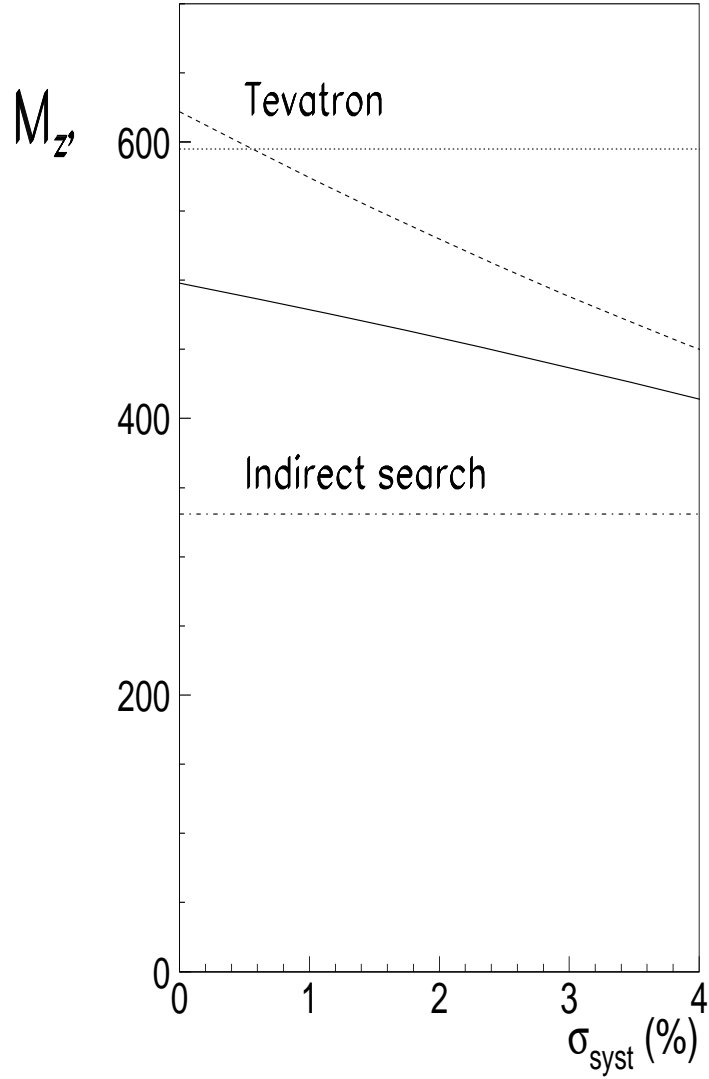,height=15cm,width=10cm}}}
\caption{Attainable sensitivity to the mass of an extra gauge boson at
95 \% C. L. in the $\chi$ model for the LAMA proposal as a function of
the systematic error per bin. The results consider the cases of a
detector of 400 kg (solid line) as well as 1 tone (dashed line).}
\label{mass}
\end{figure}

In order to estimate the sensitivity that our proposed experiment to
probe the $Z^\prime$ parameters can reach we first consider the
idealized case where the background is set to zero. The more realistic
case will be discussed later. We perform a hypothetical fit assuming
that the experiment will measure the Standard Model prediction and
adopt the same treatment as in \cite{msv}, where the reader can find a
more detailed explanation of the analysis. In order to simulate a
realistic situation we also ascribe several values for the systematic
error per bin (in per cent). The result is summarised in
Fig. \ref{mass}. One can see that, if we consider only the statistical
error one can probe at 95 \% C. L. a 500 GeV $Z^\prime$ mass in the
$\chi$ model.  This value will decrease as the systematic error
increases. In the same Fig. \ref{mass} we also show the expectations
for the case of a one tone detector. In this case the attainable
sensitivity will be 625 GeV if only the statistical error is
considered.  This value is comparable to the present constraint
obtained by the CDF collaboration.

\begin{figure}
\centerline{\protect\hbox{\psfig{file=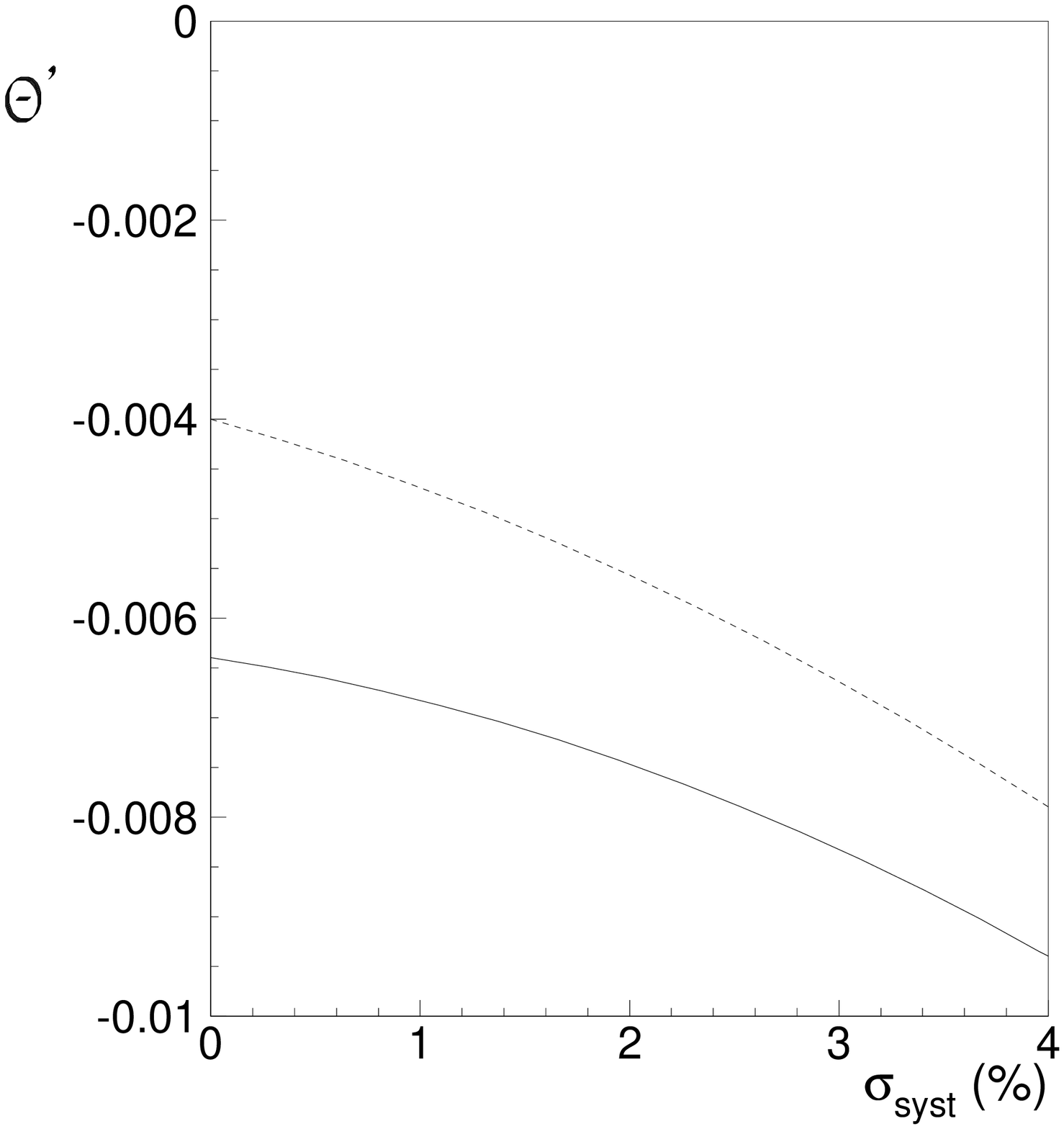,height=15cm,width=10cm}}}
\caption{Attainable  sensitivity to the mixing angle of an extra 
gauge boson in the constrained $\chi$ model for the LAMA proposal as a
function of the systematic error per bin. The results consider the
cases of a detector of 400 kg (solid line) as well as 1 tone (dashed
line).}
\label{angle}
\end{figure}

Since the sensitivity to the mixing parameter $\theta '$ is rather
poor in this kind of experiments any value of $\theta '$ in the
present allowed region will give the same result.  Therefore in our
analysis we will assume $\theta^\prime=0$ for simplicity. This is all
the model-independent information we can extract. However, having done
that, as we mentioned in the introduction, for the case of {\sl
constrained} $E_6$ models we can translate the sensitivity on
$M_{Z^\prime}$ into a corresponding sensitivity on $\theta^\prime$
using the model-dependent relationship between the mass of the extra
gauge boson $M_{Z^\prime}$ and the mixing angle $\theta^\prime$
expected in these theories.  Therefore it is possible to infer from
Fig. \ref{mass} the potential sensitivity on the mixing angle for this
specific case. This is shown in Fig \ref{angle}. One can see that the
sensitivity will be close to the LEP bound for the general
(unconstrained) model case, which applies also to the constrained
case.

\vskip 1cm
{\bf Background}

The above estimates have been made without taking into account the
background.  The required background level is determined from the
condition that statistical fluctuation of the background event number
plus standard electroweak model event number should be less than the
effect expected. Taking into account our discussion in the previous
section we conclude that the background should be less than effect
expected in the Standard Model.  So the rate of background due to
residual radioactive contaminants should be $< \: 10^{-3}$ /day/kg.
It is clear that the background requirements for this measurement are
more stringent than e.g. the ones satisfied at present by the NaI(Tl)
detectors used in the DAMA installation
\cite{Psd96,Mod97,Mod98,NCi98}.

The quantitative investigations of ref. \cite{NCi98} on the
radiopurity of these detectors and two independent preliminary
analyses \cite{Gio98} of the experimental energy spectra, from 2
keV to the MeV energy region, showed that in the relevant 2-20 keV
energy region the residual internal standard contaminants
($^{238}$U,$^{232}$Th and $^{40}$K) should have a counting rate much
lower than already measured. This result suggests that most of the
background should arise from the external environment and/or
potentially from possible internal non-standard contaminants.  In
fact, up to now low background NaI(Tl) detectors have been developed
selecting at certain level the materials to be used in the crystal
growing and taking account only the standard contaminants. Up to now
only general statements on the handling during preparation in
industrial environment have been given. Although this strategy has
so-far worked sufficiently well for the old specifications, it was
limited by the the fact that the final residual contaminations in the
detectors could be somehow different from the expected one (causing
also some differences from one detector to another); in fact, e.g.:
\begin{itemize}
\begin{enumerate}
\item 
the uniformity of the contaminants distribution inside the total
material used to construct each part of the detectors has not been
checked.
\item 
the materials were activated at sea level to some extent depending on
the period they were outside the underground site.
\item 
the uniformity of the purification effect of the crystallisation
process in the whole large mass crystalline bulk, from which more than
one detector was cut, has not been checked.
\item 
the different growing of the large crystalline bulks (needed to build
a large number of detectors) could cause different levels of residual
contaminants, because of different casual pollution and/or slight
differences in the used materials, seeds, cleaning procedures of the
crucible, etc. 
\item taking into account that the detectors are built in an industrial 
environment during several months or more (depending on the number of
detectors) one might expect possible (different) casual pollution
during the growth, the polishing and the test handling procedures
\end{enumerate}
\end{itemize}

In particular, regarding the cosmogenic activation we recall that a
great variety of long-lived radioactive isotopes are produced by
cosmic-ray spallation reactions in the detectors during the time of
detector creation. Even at the current stage of low-background
experiments, the background from these isotopes ($^{54}$Mn, $^{57}$Co,
$^{58}$Co, $^{60}$Co, $^{65}$Zn) is comparable with that arising from
other sources. They will represent certainly the main sources of
background for the next generation of experiments. The only way to
avoid their production inside the detectors is to build them deep
underground (see later).  We consider this a necessary step in
low-background detector development for the next generation
experiments.

In order to overcome the above limitations and to develop higher
radiopure detectors, suitable in particular to investigate the
$Z^\prime$ mass under consideration here, the following approach has
been studied:
\begin{itemize}
\begin{enumerate}
\item 
the first step consists in the usual careful selection of all the
required materials with the low background Germanium detector deep
underground.  This ensures the radiopurity of all materials used for
building the detectors
\item 
the second step is totally new. It consists in the selection of all
the required materials with a high sensitive mass spectrometer (MS)
and/or by neutron activation, including measurements of the more
important non-standard contaminants (see \cite{NCi98}). 
\item subsequent chemical purification of the powders, by using 
specific additives for every radioactive element. Several purification
cycles can be performed.  This purification stage has never been
utilised before and will ensure an important further purification of
the selected powders.
\item growth and assembling of the crystal deep underground in a high 
quality clean room under control of the proper operating conditions by
experimentalists. This will definitively minimise: i) the possible
casual pollution with respect to an industrial environment; ii) the
activation at sea-level of all the materials. Such program for NaI(Tl)
purification and growth has been already developed. 
\end{enumerate}
\end{itemize}

Particular care should be taken to avoid any casual pollution,
handling with extreme care the detectors deep underground. Moreover,
the detectors should never be exposed to neutron source to avoid their
activation and the activation of the surrounding materials.  Because
the detectors obviously do not measure only internal contaminants, but
also the contribution arising from the environment, the shield
materials nearest to the detectors should undergo a further complete
selection. Moreover, a new generation of low radioactive PMTs is under
consideration to reduce their significant background contribution
\cite{Gio98}; some preliminary work has already been initiated along
these lines.

Furthermore, as regards the background arising from surviving cosmic
rays deep underground, since the experiment is planned to be carried
out deep underground, the expected muon cosmic-ray intensity is
$\lsim$ 1/hour/m$^2$ which is therefore small enough for our
background requirements. Moreover it could be further decreased ---
even by 4-5 orders of magnitude --- introducing, if necessary, a
suitable anti-coincidence system.  

In summary, we conclude that the optimal region for data-taking in our
proposed experiment is 2-30 KeV. 
The statistical accuracy in this case would be only 10 \% lower than
for the full recoil electron energy range (0-100 keV) but the allowed
background level would be 3 times higher.

\section{Discussion and Conclusion.}

Some Extended Gauge Theories, such as those based on the $E_6$ group,
predict an increase of the anti-neutrino-electron scattering cross
section at low energies. For definiteness we concentrate on one of
such models, the $\chi$ model. We have proposed to look for extra
contribution of the heavy $Z^\prime$ boson in $\overline{\nu}-e$
scattering in an experiment (LAMA) with a high-activity artificial
neutrino source and with a large-mass NaI(Tl) detector at the Gran
Sasso underground laboratory. The neutrino flux is known to within a
one percent accuracy, so that even a few percent increase would be
detectable. For low enough background the sensitivity to the
$Z^\prime$ boson mass would reach 600 GeV for one year running of the
experiment.

\vspace{1cm} \noindent {\bf Acknowledgements} 

This work was supported by DGICYT under grant number PB95-1077, by the
TMR network grant ERBFMRXCT960090 and by INTAS grant 96-0659 of the
European Union. O. G. M. was supported by CONACYT and SNI-M\'exico and
V. S. by the RFFR grant 97-02-16501. I. B., V. I. G.  and
V. N. K. were supported by the Russian grant RFFI-97-02-16383.

\newpage

\begin{table}
\caption{Quantum numbers of the particles in the {\bf 27} of $E_6$.}
\begin{tabular}{cccc} \hline
  & $T_3$ &  $\sqrt{40}Y_{\chi}$ &  $\sqrt{24} Y_{\psi}$ \\
\hline
$Q$      & $\pmatrix{1/2 \cr -1/2 \cr}$  & -1   &  1  \\
         &                               &      &     \\
$u^{c}$  & 0                             & -1   &  1  \\
$e^{c}$  & 0                             & -1   &  1  \\
$d^{c}$  & 0                             &  3   &  1  \\
$l$      & $\pmatrix{1/2 \cr -1/2 \cr}$  &  3   &  1  \\
         &                               &      &     \\
$H_{d}$  & $\pmatrix{1/2 \cr -1/2 \cr}$  & -2   & -2  \\
         &                               &      &     \\
$g^{c}$  & 0                             & -2   & -2  \\
$H_{u}$  & $\pmatrix{1/2 \cr -1/2 \cr}$  &  2   & -2  \\
         &                               &      &     \\
$g$      & 0                             &  2   & -2  \\
$\nu^{c}$& 0                             & -5   &  1  \\
$n$      & 0                             &  0   &  4  \\
\hline
\end{tabular}
\end{table}

\begin{table}
\caption{The expected number of events for the LAMA proposal for the Standard 
Model and for the $\chi$ model for different values of the $Z^\prime$ mass. }
\begin{tabular}{ccccccc}
\hline 
{Electron recoil} &          &         &       &       &       &     \\
Energy                &0-20 keV  & 20-40   &40-60  & 60-80 & 80-100&Total \\
\hline
St. Model             & 8472     & 5207    & 2899   & 1359   & 454 & 18391 \\
\hline
Ext. Model            & 8790     & 5402    & 3007   & 1409   & 470 & 19078 \\
(330 GeV)             & & & & & & \\
Difference            &  318     &  195    &  108   &   50   &  16 &   687 \\
\hline
Ext. Model            & 8570     & 5267    & 2931   & 1374   & 458 & 18600 \\
(600 GeV)             & & & & & &  \\
Difference            &   98     &   60    &   32   &   15   &   4 &   209 \\
\hline
Ext. Model            & 8543     & 5251    & 2923   & 1370   & 457 & 18544 \\
(700 GeV)             & & & & & &  \\
Difference            &   71     &   44    &   24   &   11   &   3 &   153 \\
\hline
Ext. Model            & 8507     & 5228    & 2910   & 1365   & 455 & 18465 \\
(1000 GeV)            & & & & & &  \\
Difference            &   35     &   21    &   12   &    6   &   1 &    75 \\
\hline
\end{tabular}
\end{table}

\begin{thebibliography}{9}

\bibitem{Barabanov} 
I. R. Barabanov {\it et. al.},  {\it Astrop. Phys.} {\bf 8} (1997) 67; 
I. R. Barabanov {\it et. al.}, {\it Astrop. Phys.} {\bf 5} (1996) 159.

\bibitem{Vogel}
P. Vogel and J. Engel, {\it Phys. Rev. }{\bf D39 } (1989) 3378.

\bibitem{Ferrari} 
N. Ferrari, G. Fiorentini and B. Ricci {\it Phys. Lett.} {\bf B387} 
(1996) 427.

\bibitem{msv}
O. G. Miranda, V. Semikoz and  Jos\'e W. F. Valle 
hep-ph/9712215, {\it Phys. Rev.} {\bf D 58} (1998) 013007.

\bibitem{Broggini} 
M. Moretti, C. Broggini and G. Fiorentini, {\it Phys. Rev.} {\bf D 58}
(1998) 4160.

\bibitem{fae}
See, for example, J.W.F. Valle, 
{\it Prog. Part. Nucl. Phys. }{\bf 26} (1991) 91, and references
therein.

\bibitem{GCprd} M. C. Gonz\'alez-Garc\'{\i}a and J. W. F. Valle, 
{\it Phys. Rev.} {\bf D41} (1990) 2355.

\bibitem{npb345}
M. C. Gonz\'alez-Garc\'{\i}a and J. W. F. Valle, 
{\it Nucl. Phys. }{\bf B345} (1990) 312.

\bibitem{LR1}
J.C. Pati and A. Salam, 
{\it Phys. Rev. }{\bf D10 } (1975) 275;
R.N. Mohapatra and J.C. Pati, {\it Phys. Rev. }{\bf D11 } (1975) 566, 2558.

\bibitem{LR2}
R.N. Mohapatra and G. Senjanovi\'c, 
{\it Phys. Rev. }{\bf D23 } (1981) 165
and references therein. 

\bibitem{lang}
{\sl Precision Tests of the Standard Electroweak Model}.
P. Langacker, (ed.), Singapore, Singapore: World Scientific (1995)
(Advanced series on directions in high energy physics: 14);
M. Cvetic, P. Langacker, hep-ph/9707451.

\bibitem{CDF}
CDF coll., F. Abe et. al., {\it Phys. Rev. Lett.} {\bf 79} (1997) 2192.

\bibitem{Fayans}
S. A. Fayans, V. Yu. Dobretsov, A. B. Dobrotsvetov, {\it Phys. Lett.} 
{\bf B291} (1992) 1.

\bibitem{Fayans2}
V. I. Kopeikin, L. A. Mikaelyan, V. V. Sinev and S. A. Fayans, 
{\it Phys. of Atomic Nuc.} {\bf 60} (1997) 2032.

\bibitem{Fayans3}
S. A. Fayans, L. A. Mikaelyan, V. V. Sinev, {\sl Inelastic scattering
of antineutrinos from reactor and beta sources on atomic electrons},
Wein-98, Santa Fe, June 1998.

\bibitem{Kornoukhov}
V.N.Kornoukhov, Preprint ITEP N90 (1994), ITEP N26 (1996), ITEP N2 (1997)
and {\it Phys. of Atomic Nuc.} {\bf 60} (1997) 558.

\bibitem{Bernabei}
R. Bernabei {\it et. al.} {\it Phys. Lett.} {\bf B389} (1996)  757, 
R. Bernabei {\it et al.}  {\it Phys. Lett.} {\bf B424} (1998) 195.

\bibitem{NCi98} R. Bernabei et al., ROM2F/98/27 August 1998.
\bibitem{Psd96} R. Bernabei et al., {\it Phys. Lett.} {\bf B389} (1996), 757. 
\bibitem{Mod97} R. Bernabei et al., {\it Phys. Lett.} {\bf B424} (1998), 195.
\bibitem{Mod98} R. Bernabei et al., ROM2F/98/34, 
  August 1998 and INFN/AE-98/20.  
\bibitem{Gio98} G. Ignesti, thesis, Universita' di Roma 
 "La Sapienza", oct.1998.

\end{thebibliography}
\end{document}